\documentclass[twocolumn]{aastex61}
\pdfoutput=1

\newcommand{\um}{\,\micron}

\newcommand{\HII}{H\,\textsc{ii}}
\newcommand{\cII}{[\textrm{C}\,\textsc{ii}]}
\newcommand{\cIItir}{[\textrm{C}\,\textsc{ii}]/\textrm{TIR}}

\newcommand{\oI}{[O\,\textsc{i}]}

\newcommand{\nII}{[N\,\textsc{ii}]}
\newcommand{\oh}{12+log(\textrm{O}/\textrm{H})}
\newcommand{\kf}{{\sc Kingfish}}
\newcommand{\sfrd}{\ensuremath{\Sigma_\textsc{sfr}}}
\newcommand{\sfdunt}{\ensuremath{\mathrm{M}_\sun/\mathrm{yr}/\mathrm{kpc}^2}}

\newcommand{\nuInu}[1]{\ensuremath{\nu I_\nu(#1)}}

\shortauthors{Smith \textit{et al.}}  
\shorttitle{Resolving the Cooling Line Deficit}
\begin{document}

\title{The Spatially Resolved \cII\ Cooling Line Deficit in Galaxies}
\author{J.D.T. Smith}	%
\affiliation{Ritter Astrophysical Research Center, University of Toledo, Toledo, OH 43606, USA}
\affiliation{Max-Planck-Institut f\"{u}r Astronomie, K\"{o}nigstuhl 17, D-69117 Heidelberg, Germany} %
\email{jd.smith@utoledo.edu}
\author{Kevin Croxall}	%
\affiliation{Department of Astronomy, The Ohio State University, 140 W 18th Ave., Columbus, OH 43210, USA}%
\affiliation{Illumination Works LLC, 5650 Blazer Parkway, Suite 152, Dublin, OH 43017}
\author{Bruce Draine}	%
\affiliation{Department of Astrophysical Sciences, Princeton University, Princeton, NJ 08544, USA}%
\author{Ilse De Looze}	%
\affiliation{Institute of Astronomy, University of Cambridge, Madingley Road, Cambridge, CB3 0HA, UK}%
\author{Karin Sandstrom}	%
\affiliation{Center for Astrophysics and Space Sciences, University of California, San Diego, 9500 Gilman Drive, La Jolla, CA 92093, USA}
\author{Lee Armus}	%
\affiliation{Spitzer Science Center, California Institute of Technology, MC 314-6, Pasadena, CA 91125, USA}%
\author{Pedro Beir\~{a}o}	%
\affiliation{Observatoire de Paris, 61 avenue de l’Observatoire, F-75014 Paris, France}%
\author{Alberto Bolatto}	%
\affiliation{Department of Astronomy, University of Maryland, College Park, MD 20742, USA}%
\author{Mederic Boquien}	%
\affiliation{Unidad de Astronomía, Fac. Cs. Básicas, Universidad de Antofagasta, Avda. U. de Antofagasta 02800, Antofagasta, Chile}
\author{Bernhard Brandl}	%
\affiliation{Leiden Observatory, Leiden University, P.O. Box 9513, 2300 RA Leiden, The Netherlands}
\affiliation{Faculty of Aerospace Engineering, Delft University of Technology, Kluyverweg 1, 2629 HS Delft, The Netherlands}
\author{Alison Crocker}	%
\affiliation{Department of Physics, Reed College, Portland, OR 97202, USA}%
\author{Daniel A. Dale}	%
\affiliation{Department of Physics and Astronomy, University of Wyoming, Laramie, WY 82071, USA}%
\author{Maud Galametz}	%
\affiliation{European Southern Observatory, Karl-Schwarzschild-Str. 2, 85748 Garching-bei-München, Germany}%
\author{Brent Groves}	%
\affiliation{Research School of Astronomy \& Astrophysics, Australian National University, Canberra, ACT 2611, Australia}
\author{George Helou}	%
\affiliation{Infrared Processing and Analysis Center, California Institute of Technology, 1200 E. California Boulevard, Pasadena, CA 91125, USA}
\author{Rodrigo Herrera-Camus}	%
\affiliation{Max Planck Institute for Extraterrestrial Physics Garching, Germany}
\author{Leslie Hunt}	%
\affiliation{INAF-Osservatorio Astrofisico di Arcetri, Largo E. Fermi 5, 50125 Firenze, Italy} %
\author{Robert Kennicutt}	%
\affiliation{Institute of Astronomy, University of Cambridge, Madingley Road, Cambridge, CB3 0HA, UK}%
\author{Fabian Walter}	%
\affiliation{Max-Planck-Institut f\"{u}r Astronomie, K\"{o}nigstuhl 17, D-69117 Heidelberg, Germany} %
\author{Mark Wolfire}   %
\affiliation{Department of Astronomy, University of Maryland, College Park, MD 20742, USA}%

\begin{abstract}
We present \cII\ 158\um\ measurements from over 15,000 resolved regions within 54 nearby galaxies of the \kf\ program to investigate the so-called \cII\ ``line cooling deficit'' long known to occur in galaxies with different luminosities.  The \cIItir\ ratio ranges from above 1\% to below 0.1\% in the sample, with a mean value of $0.48\pm0.21$\%.  We find that the surface density of 24\um\ emission dominates this trend, with \cIItir\ dropping as \nuInu{24\um} increases.  Deviations from this overall decline are correlated with changes in the gas phase metal abundance, with higher metallicity associated with deeper deficits at a fixed surface brightness.  We supplement the local sample with resolved \cII\ measurements from nearby luminous infrared galaxies and high redshift sources from z=1.8--6.4, and find that star formation rate density drives a continuous trend of deepening \cII\ deficit across six orders of magnitude in \sfrd.  The tightness of this correlation suggests that an approximate \sfrd\ can be estimated directly from global measurements of \cIItir, and a relation is provided to do so.  Several low-luminosity AGN hosts in the sample show additional and significant central suppression of \cIItir, but these deficit enhancements occur not in those AGN with the highest X-ray luminosities, but instead those with the highest central starlight intensities.  Taken together, these results demonstrate that the \cII\ cooling line deficit in galaxies likely arises from local physical phenomena in interstellar gas.

\end{abstract}


\section{Introduction}

The flow of radiative energy through the gas and dust comprising the interstellar medium (ISM) of galaxies strongly impacts their observational and physical properties.  Indeed, the process of star formation itself is controlled in part by the balance between the radiative heating and cooling in star-forming clouds of gas and dust \citep[e.g.][]{Krumholz2011}.  With high ultraviolet (UV) opacity, dust grains absorb and re-radiate a significant fraction of a galaxy's radiative energy \citep[from a few percent up to nearly 100 percent in the most infrared-luminous objects,][]{Sanders1996}.  The smallest of these dust grains have small heat capacities, and so are stochastically heated to high non-equilibrium temperatures \citep{Draine2001a}.  With their modest ionization energies \citep[$\sim$6--9eV,][]{Witt2006} and large cross-sections, these small grains, including polycyclic aromatic hydrocarbons (PAHs), are also the dominant source of the photoelectrons which energetically couple the radiation field to the gas outside of hydrogen-ionized zones \citep{Bakes1994}.  

Neutral gas cools predominantly through collisional excitation of abundant ions with low-lying excited states.  With a low first ionization potential (11.26\,eV) well below the Lyman limit, atomic carbon is found in singly-ionized form in a wide range of environments.  Together with its high abundance and modest first excitation energy, this makes ionized carbon a highly effective coolant of neutral atomic and even diffuse molecular gas.  Indeed, for this reason, the most luminous emission line of a galaxy is typically \cII\ 158\um\  \citep[e.g.][]{Stacey1991}.  This high line luminosity (up to several percent of the total infrared output) also makes \cII\ a promising diagnostic tool for mapping the key physical properties of galaxies in the early universe, including star formation rate, kinematics, size, virial mass and molecular content.  One of ALMA's primary science drivers involves imaging \cII\ in Milky Way type galaxies at redshifts greater than 3, and interferometers like IRAM/PdBI and ALMA are now detecting \cII\ in galaxies beyond redshift z=6--7 \citep[e.g.][]{Venemans2012,Capak2015,Maiolino2015,Willott2015}.

\citet{Helou2001} used the near constancy of \cII\ relative to polycyclic aromatic hydrocarbon (PAH) emission in support of models that stressed the importance of very small dust grains to photo-electric heating of neutral gas. But for stellar populations drawn from realistic star formation histories, the small grains which heat gas via photo-electrons, and the larger grains which re-process the bulk of absorbed UV/Optical radiative energy into the infrared do not differ substantially in their contributions to absorption of starlight \citep[e.g.][]{2013ApJ...762...79C}.  Thus, the rate of gas cooling via \cII\ might be expected to closely track the total infrared (TIR, 3--1100\um) emission of galaxies --- \cIItir\ should remain approximately constant.  Indeed the suitability of \cII\ as a direct star-formation tracer, which follows from this expectation, has been given considerable attention \citep{Boselli2002,Farrah2013,DeLooze2014,Kapala2015,Herrera-Camus2015}.

And yet, relative to the heating rate inferred from bolometric infrared luminosity, the line luminosity of this principal coolant is found to vary by \emph{orders of magnitude} among different galaxy types and environments \citep{Malhotra1997,Malhotra2001,Luhman2003,Brauher2008,Stacey2010,Gracia-Carpio2011,Diaz-Santos2013}.  
This so-called \cII\ \emph{cooling line deficit} problem, first revealed as an unexpectedly low \cII\ luminosity among local Ultra-Luminosity Infared Galaxies (ULIRGs), has proven difficult to reconcile, with possible explanations invoking physical impacts on either the coupling of radiation to the gas via dust grains, or on the nature of the subsequent gas cooling itself.

The Herschel Space Observatory has for the first time delivered the combination of sensitivity and spatial resolution necessary to map the fundamental cooling lines on sub-kiloparsec scales in nearby galaxies.  \kf\ \citep[Key Insights on Nearby Galaxies --- a Far-Infrared Survey with \textit{Herschel};][]{Kennicutt2011} has investigated a diverse sample of galaxies in the nearby universe (d$\lesssim$30\,Mpc).  Here we explore resolved maps of \cII\ across the \kf\ sample in order to delineate the principal physical drivers of the deficit itself.  In \S\,\ref{sec:observ-reduct} we detail the \textit{Herschel} and supporting observations and data reduction and demonstrate the resulting quality of the \cII\ maps; in \S\,\ref{sec:type} we highlight the impact of surface brightness and metallicity on the deficit; in \S\,\ref{sec:sfrd} we evaluate the role played by star formation rate density, both locally in normal and luminous infrared galaxies, and at high redshift; in \S\,\ref{sec:agn-impact} we investigate the spatially resolved impact of active galactic nuclei (AGN), concluding with discussions in \S\,\ref{sec:discussion} and \S\,\ref{sec:conclusions}.

\section{Observations and Data Reduction}
\label{sec:observ-reduct}

\subsection{PACS Observations}
\textit{Herschel}/PACS was used to perform far-IR mapped spectral observations of the \cII\ 158$\mu$m line as part of the \textit{Herschel} Open Time Key Program \kf.  Observations targeted radial strips centered on galactic nuclei as well as additional extranuclear regions.   The targeted galaxies cover a wide range of star formation rate (0.02--11$\:\mathrm{M}_\sun/\mathrm{yr}$), infrared-to-optical luminosity ratio (0.02--6.2), and total gas mass ($10^{7.8}$--$10^{10.3}\:\mathrm{M}_\sun$). For more information on the target selection, including the mapped regions within sample galaxies, see \citet{Kennicutt2011}.

All PACS spectral observations were obtained in the Un-Chopped Mapping mode and reduced using the \textit{Herschel} Interactive Processing Environment (HIPE) version 11.2637.  Reductions applied the standard spectral response functions and flat field corrections, flagged instrument artifacts and bad pixels, and subtracted dark current.  Transients caused by thermal instabilities were removed through custom treatment designed for the \kf\ Pipeline.  Specific information on data reduction is contained in \citet{2012ApJ...747...81C} and the \kf\ Data Products Delivery (DR3) User’s Guide.\footnote{\url{ftp://hsa.esac.esa.int/LEGACY\_PRODUCTS/UPDP/KINGFISH-DR3/KINGFISH\_DR3.pdf}}  Flux maps were obtained by fitting single Gaussian profiles to each projected pixel, with care taken to avoid spurious detection of spectral artifacts.  When Gaussian fitting failed (for example due to appreciable velocity broadening within the region), a straight integration was adopted.  Line fits employed iterative velocity tuning, initialized with velocities from region-matched velocity measurements of atomic hydrogen from THINGS \citep{Walter2008a}, where available. Flux calibration of PACS data yield absolute flux uncertainties of $\sim$15\% with relative flux uncertainties between each \textit{Herschel} pointing within a galaxy of $\sim$10\%.

Circular extraction regions of 11\arcsec\ diameter were tiled over the available fully-sampled \cII\ mapping area in the 54 \cII-mapped \kf\ galaxies.  This extraction size is well matched to the resolution PACS delivers at 158\um, as well as that of the convolved ancillary data sets (see below).  After masking for unreliable line fits and cutting at S/N=4, approximately 15,000 distinct regions with \cII\ detection remained, ranging in physical diameter from 0.17--1.6\,kpc (median 0.65\,kpc).  Of the regions, 88\% have \cII\ detected with S/N$>$5, 60\% have S/N$>$10, and 33\% have S/N$>$20. One galaxy characterized by deep deficits --- the nascent starburst or embedded AGN host galaxy NGC\thinspace1377 \citep{2006ApJ...646..841R,Aalto2016}--- was omitted from the sample, as it has extreme attenuation conditions and an ambiguous central power source \citep[see also][]{Herrera-Camus2015}.  

\subsection{Ancillary Data}

Additional \textit{Spitzer}/SINGS and \kf\ photometric maps were used to assess the total infrared power, all convolved to match the PACS 160\um\ channel resolution \citep{2012ApJ...756..138A}.  GALEX far-ultraviolet (FUV; 0.155\um) maps, used for accurate star formation measurements at low surface brightness, were similarly convolved.  Maps of luminous infrared galaxies resolved at 24\um\ from GOALS \citep{Armus2009} were convolved to match the 160\um\ resolution using the convolution kernel methodology of \citet{Aniano2011}.  To minimize any possible effects of starlight contamination at shorter wavelengths \citep[see, e.g.][]{Kapala2015}, the calibration of total infrared (TIR, 3--1100\um) luminosity was performed utilizing 70, 100, and 160\um\ photometry, which closely tracks full-SED integrated TIR luminosity \citep{Galametz2013}.

\subsection{Resulting Line Maps}
\label{sec:results}

\begin{figure*}[t]
\centering
\begin{minipage}[b]{0.755\linewidth}
\centering
  \includegraphics[width=\textwidth]{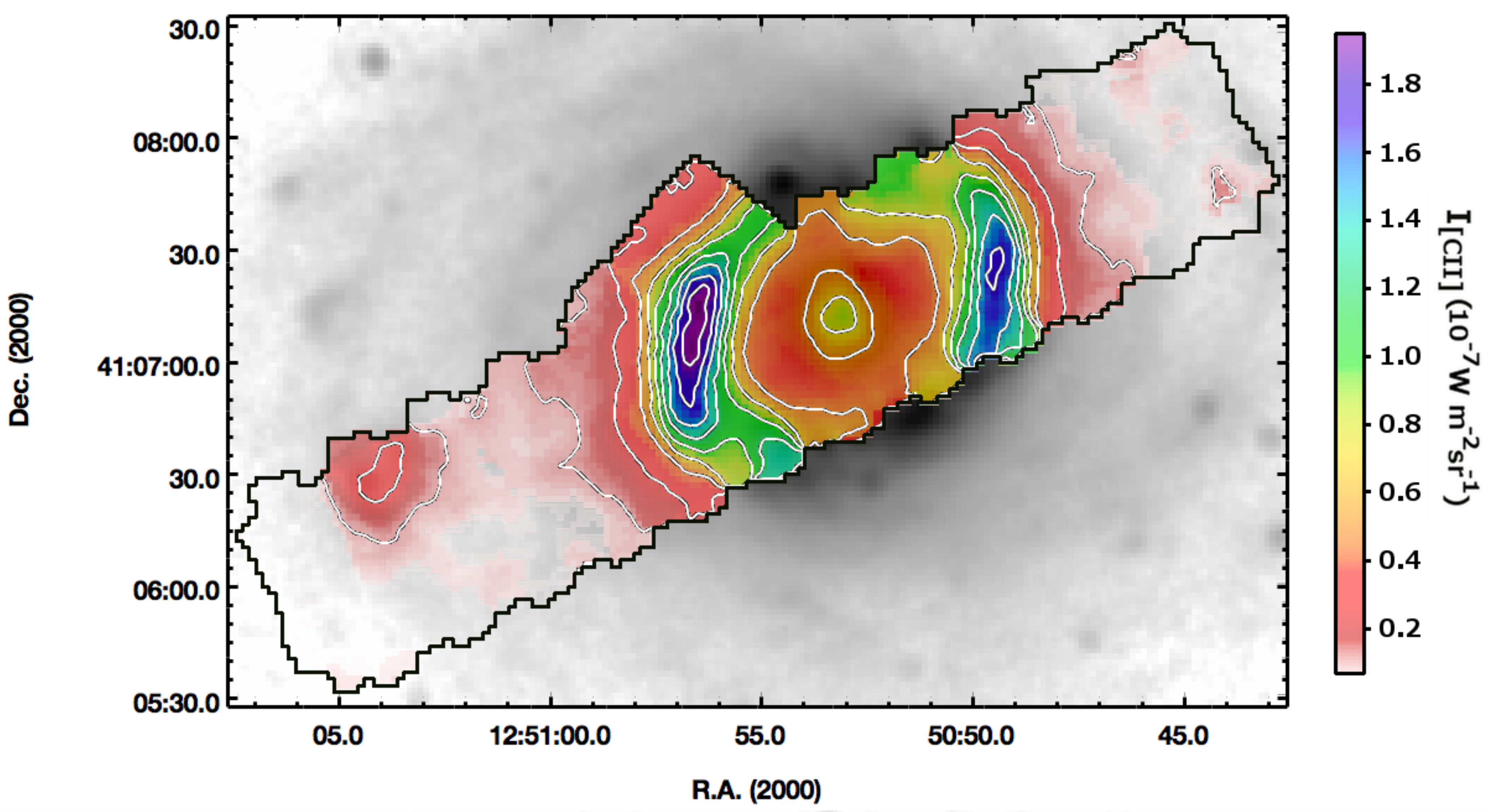}\\
  \includegraphics[width=\textwidth]{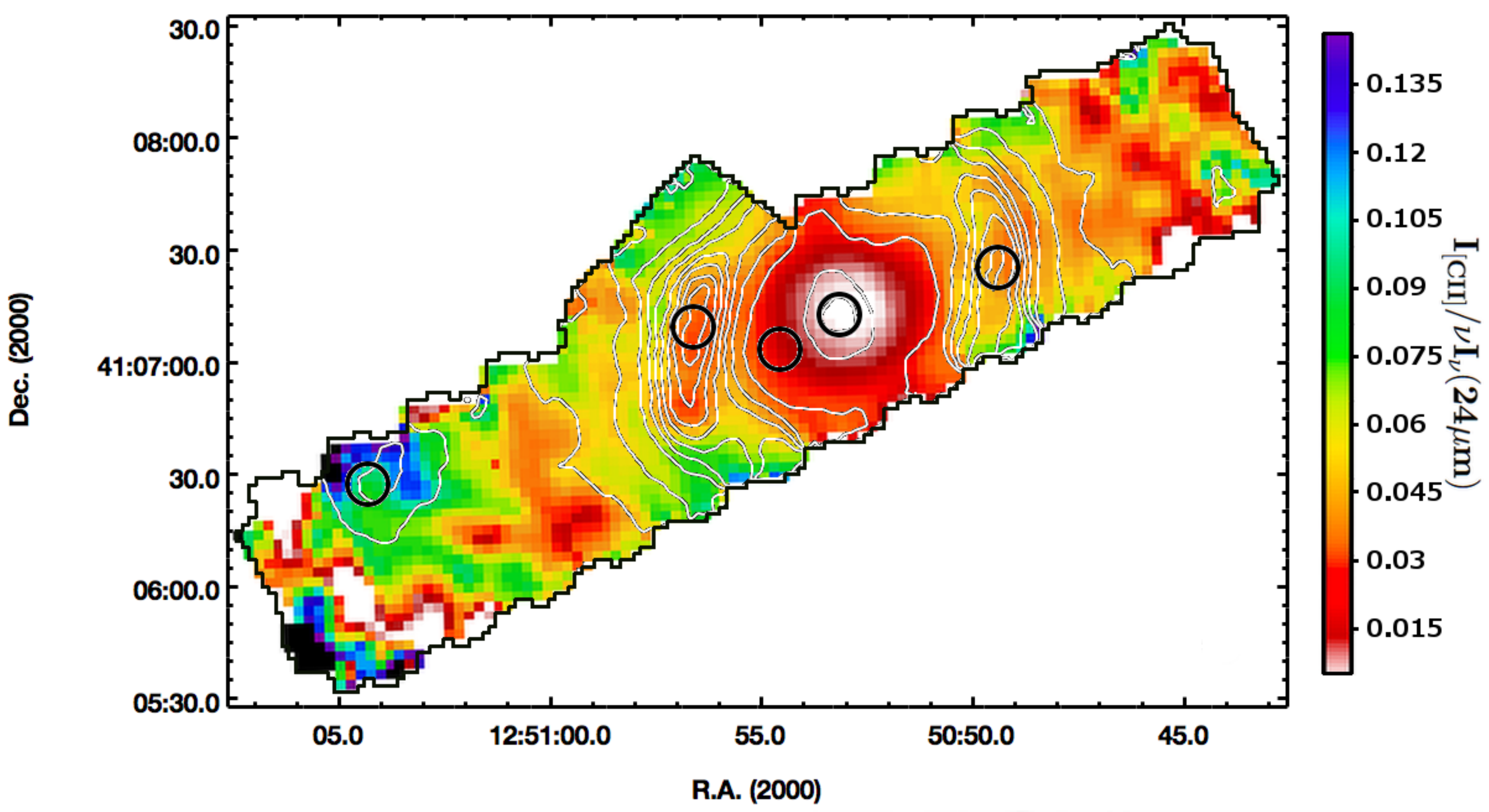}
\end{minipage}
\hspace{-.05in}
\begin{minipage}[b]{0.235\linewidth}
\centering
  \includegraphics[width=\textwidth,trim=13 11 23 24,clip=true]{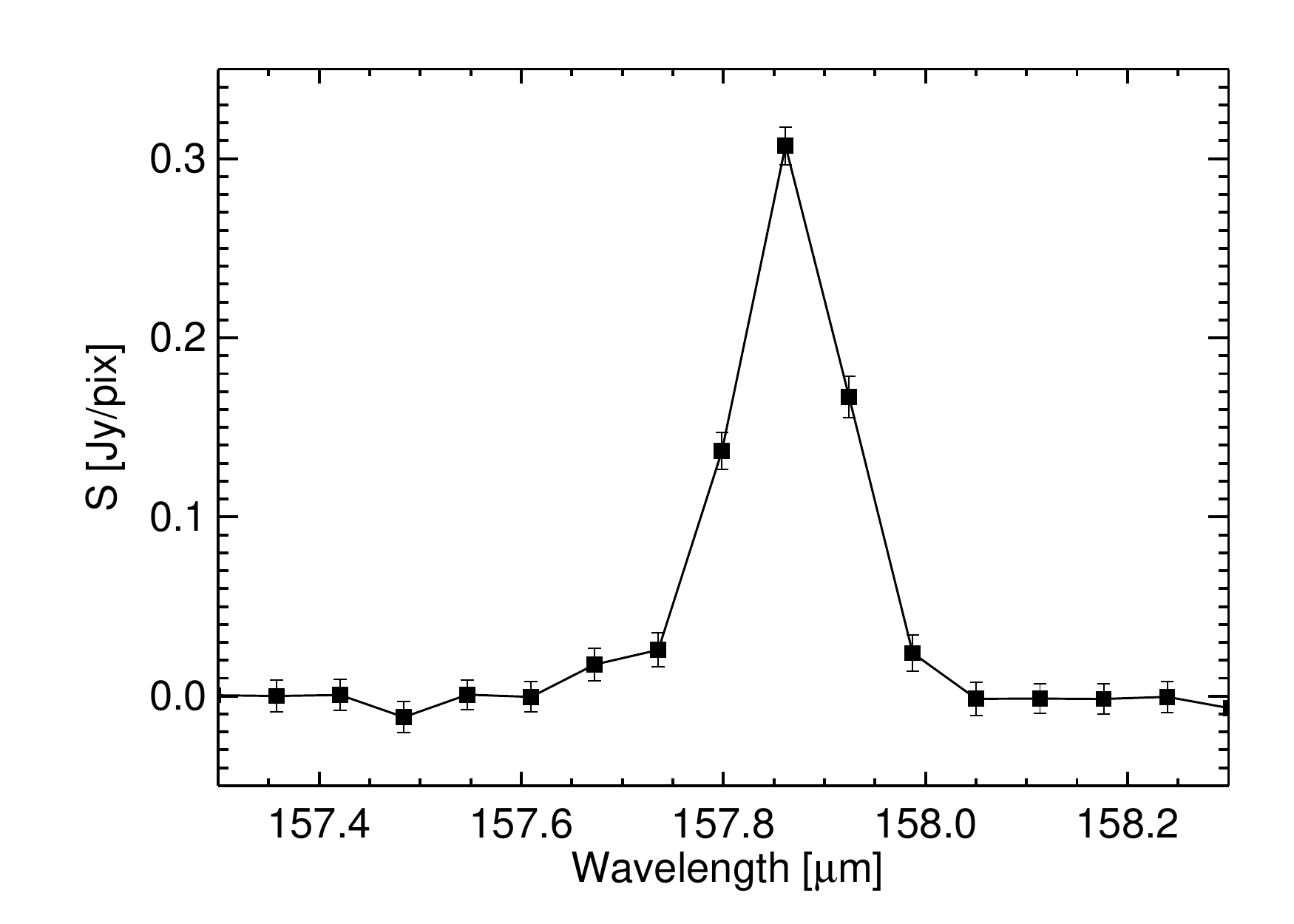}
  \includegraphics[width=\textwidth,trim=13 11 23 24,clip=true]{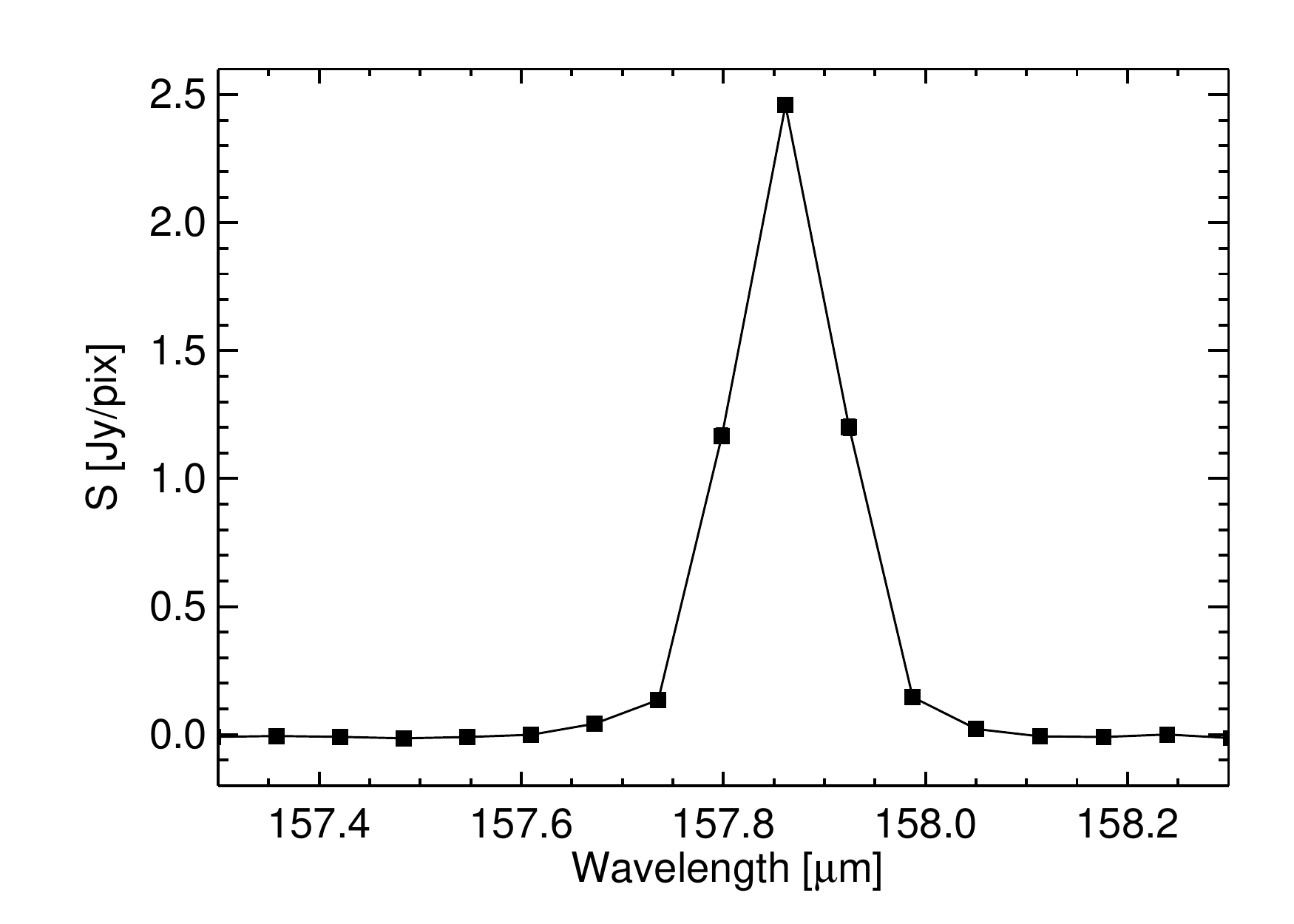}
  \includegraphics[width=\textwidth,trim=13 11 23 24,clip=true]{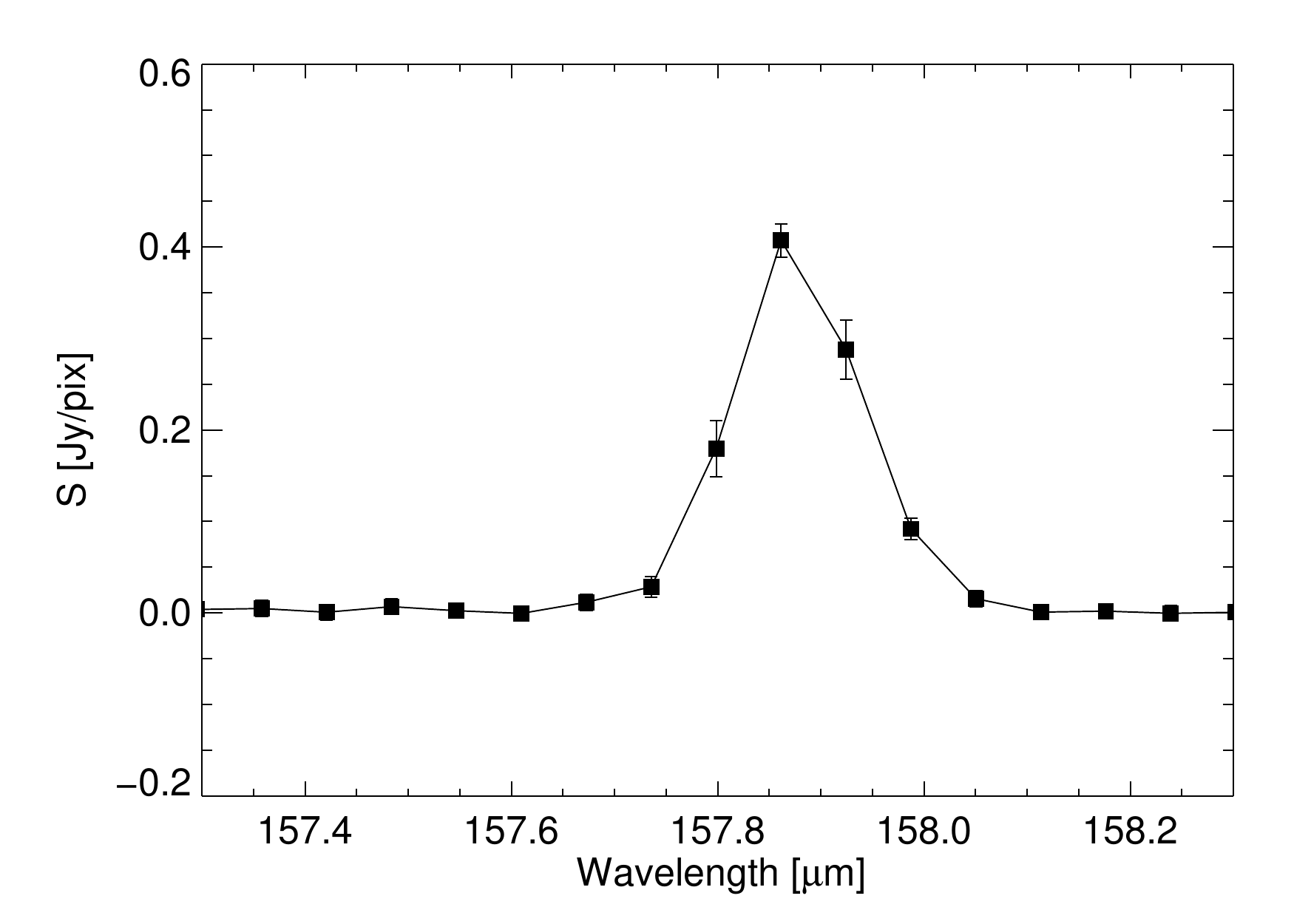}
  \includegraphics[width=\textwidth,trim=13 11 23 24,clip=true]{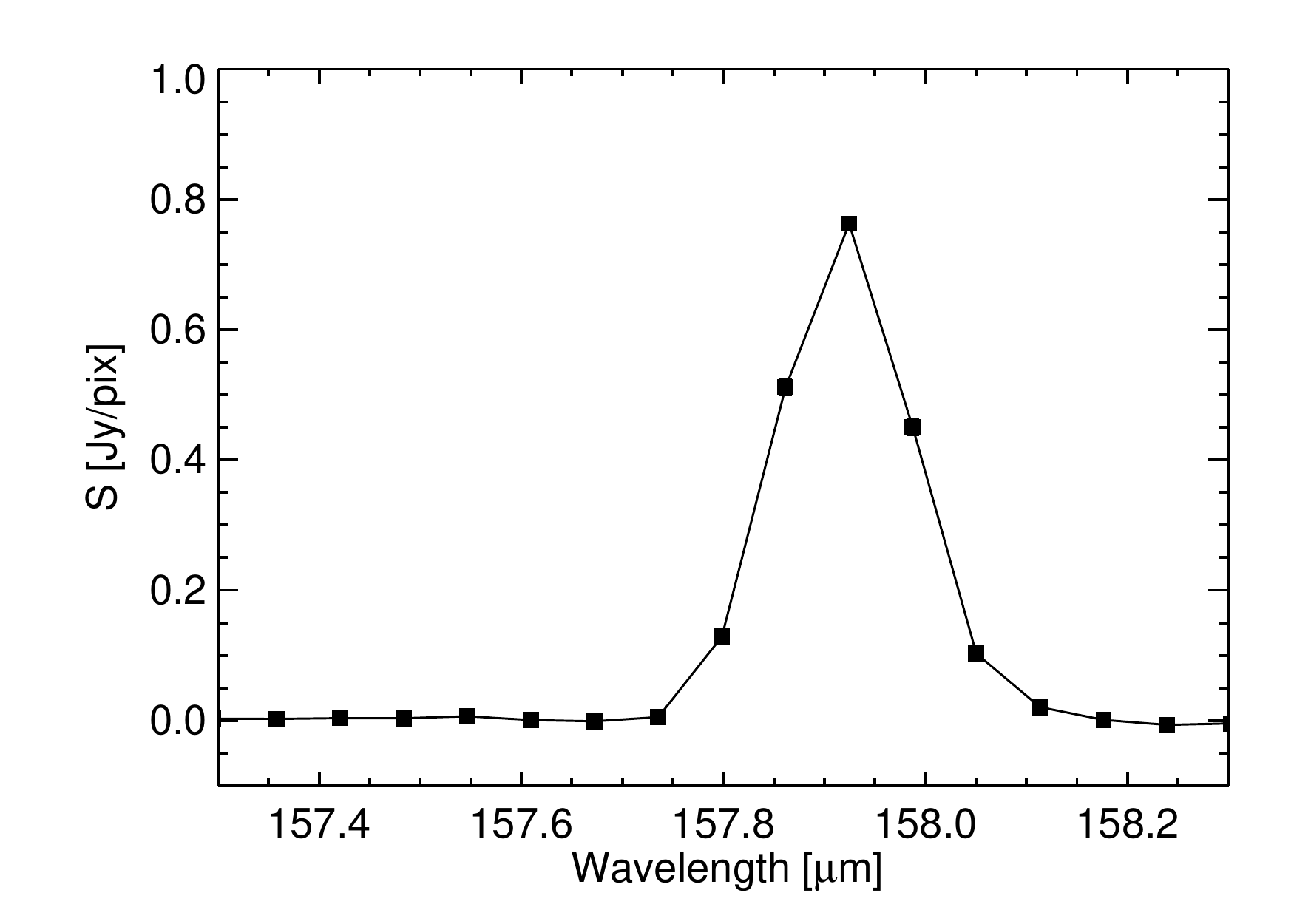}
  \includegraphics[width=\textwidth,trim=13 11 23 24,clip=true]{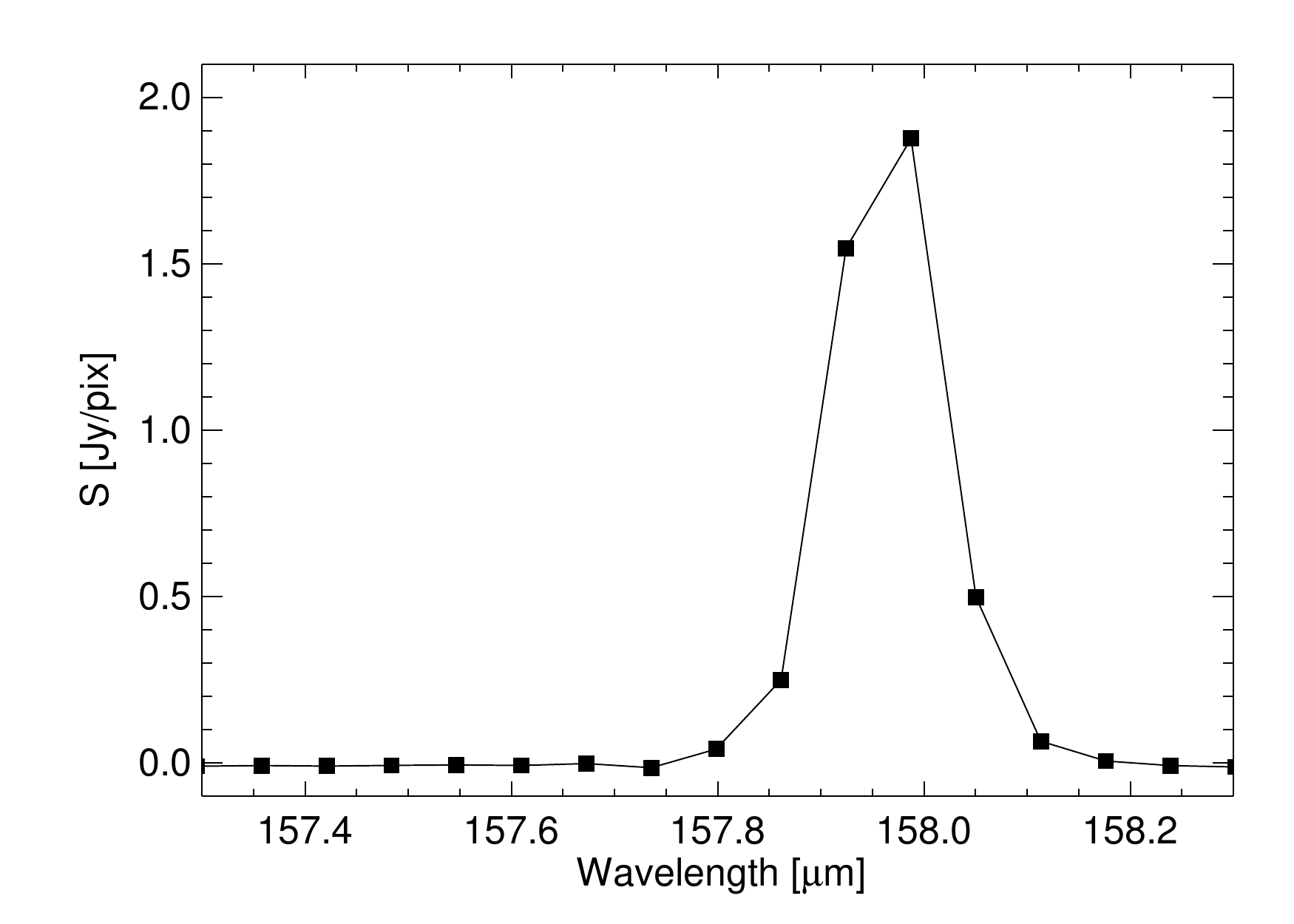}
\end{minipage}

\caption{\label{fig:cii_im} An example of the \cII\ line maps produced from \textit{Herschel}/PACS spectral mapping in \kf.  Above and left, the \cII\ contours and intensity-matched color scale are overlaid on a \textit{Spitzer}/MIPS 24\um\ image of \kf\ galaxy NGC\thinspace4736.  Coverage is along radial strips, supplemented by selected extra-nuclear targets (in this example extending the strip along the star-forming ring to the north).  Below it, the $I_{[\mathrm{CII}]}$/\nuInu{24\um} ratio map (colors, with $I_{[\mathrm{CII}]}$ contours) demonstrates the widely varying fractional \cII\ intensity.  Black circles indicate locations of five example \cII\ extraction regions, each 11\arcsec\ in diameter.  The corresponding \cII\ spectra from regions left to right are arranged from top to bottom, covering a faint outer disk region, inner arm regions, the bright star-forming ring, and the nucleus.}
\end{figure*}

The high sensitivity realized in \textit{Herschel}/PACS unchopped mapping observations translates to significant \cII\ detection in most map pixels.  An example of the \cII\ maps covering radial strips and selected extra-nuclear regions of 54 \kf\ galaxies is given in Fig.~\ref{fig:cii_im} for starburst-ring and low-luminosity AGN host galaxy NGC\thinspace4736 \citep[see also][]{2015A&A...575A..83V}.  Strong variation of \cII\ brightness relative to the underlying 24\um\ emission map is seen in the $I_{[\mathrm{CII}]}$/\nuInu{24\um} map in the lower panel.  Example extracted \cII\ line profiles from bright, intermediate, and faint areas of the spectral cube are included. In most cases, the lines are unresolved at the \textit{Herschel}/PACS spectral resolution of $\sim$239\,km/s at the \cII\ line \citep[see][]{deBlok2016}.

A complete compendium of \kf\ line fluxes, spectral cubes, and related data is under preparation. \citet{Herrera-Camus2015} used the \kf\ \cII\ data set to provide updated star formation rate calibrations of the line (see \S\,\ref{sec:sfrd}).  Here we investigate variations in \cIItir\ with important \emph{intensive} (i.e. local, not global) physical properties of galaxies, including surface brightness, gas phase metallicity, star formation rate density, and AGN proximity.

\section{The Resolved Line Deficit in Nearby Galaxies}
\label{sec:type}

While early surveys uncovered an apparent dependence of the fractional \cII\ line power on galaxies' bulk infrared luminosity \citep[e.g.][]{Malhotra2001,Luhman2003}, with the highest luminosity galaxies exhibiting the deepest deficits, the advent of \cII\ surveys at redshifts z$\gtrsim$1 has completely altered this viewpoint.  
In the nearby universe, to good approximation, global luminosity serves as a reliable parameterization of the strongly varying local heating conditions in the ISM which must underly the deficit, perhaps due to strong correlation between merger activity and high infrared luminosity.  But this is emphatically not the case at earlier epochs, where wide ranges of ISM conditions exist even at the highest infrared luminosities.   This is well summarized by \citet{Brisbin2015}, who compile local and recent higher redshift results to demonstrate conclusively that the global infrared luminosity of a galaxy is a very poor predictor of its deficit.  For example, at far-infrared (FIR, 40--122\um) luminosities $L_\mathrm{FIR}> 10^{12.5} L_\sun$, \cII/FIR is \emph{not} uniformly low as is found in local ULIRGS, but instead ranges over more than two orders of magnitude, and can in fact \emph{exceed} the value seen in local star-forming systems more than a thousand times less luminous. 
This fits directly with the developing evolutionary framework in which the highest luminosity systems at earlier epochs arise not exclusively from mergers but increasingly in large-scale, cooler extended starbursts driven by the direct accretion of cold gas from the richer available reservoirs \citep[e.g.][]{Pope2008,Menendez-Delmestre2009,Symeonidis2013}.  
Results from the ISO satellite provided another viewpoint --- that deficits are most strongly correlated with the infrared dust color temperature --- an intrinsic indirect measure of the intensity of the interstellar radiation field \citep[e.g. $I(60\um)/I(100\um)$,][]{Malhotra2001,Brauher2008}. Other recent results have extended the breakdown of infrared luminosity as a key driver of the cooling deficit to redshifts as low as $z\sim 0.2$ \citep{Rigopoulou2014,Magdis2014,Ferkinhoff2014,Brisbin2015,Ibar2015}.  And some have abandoned $L_{IR}$ as a driving parameter in favor of intrinsic quantities such as luminosity per unit gas mass \citep[which relates to the efficiency of star formation,][]{Gracia-Carpio2011}.  This leaves open the question of which physical parameters of the star-forming ISM dominate the strong observed deficit trends.  Spatially resolved samples with detailed gas and dust diagnostics on sub-kpc scales are ideally suited to exploring this.  Here we consider two alternative galaxy characteristics which together explain the bulk of the variation in \cIItir\ found in local galaxies: infrared surface brightness (which relates to the surface density of star formation), and gas phase metallicity, \oh.

\subsection{Surface Brightness}
\label{sec:surface-brightness}

A number of studies have tracked the \cII\ deficit on sub-kpc scales in galaxies. \citet{Nakagawa1998} reported balloon-borne \cII\ observations of a large portion of the Galactic plane, finding \cII/FIR decreases in star-forming regions and the Galactic center. 
\citet{2012ApJ...751..144B} found deficit variations in starburst-ring \kf\ galaxy NGC\thinspace1097, and \citet{2012ApJ...747...81C} expanded this work to an additional galaxy, NGC\thinspace4559, uncovering changes in the ratio of \cII\ to PAH emission correlated with the charge state of PAH grains.  \citet{Kramer2013} and \citet{Kapala2015} found that the fractional luminosity of \cII\ drops also towards the centers of M33 and M31, respectively.  On larger, but still resolved scales, \citet{Diaz-Santos2014} found that the deficit in the luminous infrared galaxies (LIRGs) of the GOALS sample was mostly confined to their starbursting nuclei, and that infrared surface brightness anti-correlated with \cII/FIR.  \citet{Lutz2015} extended this to quasar host galaxies and also found a tight trend of \cII/FIR with FIR surface brightness.

\begin{figure*}
  \plotone{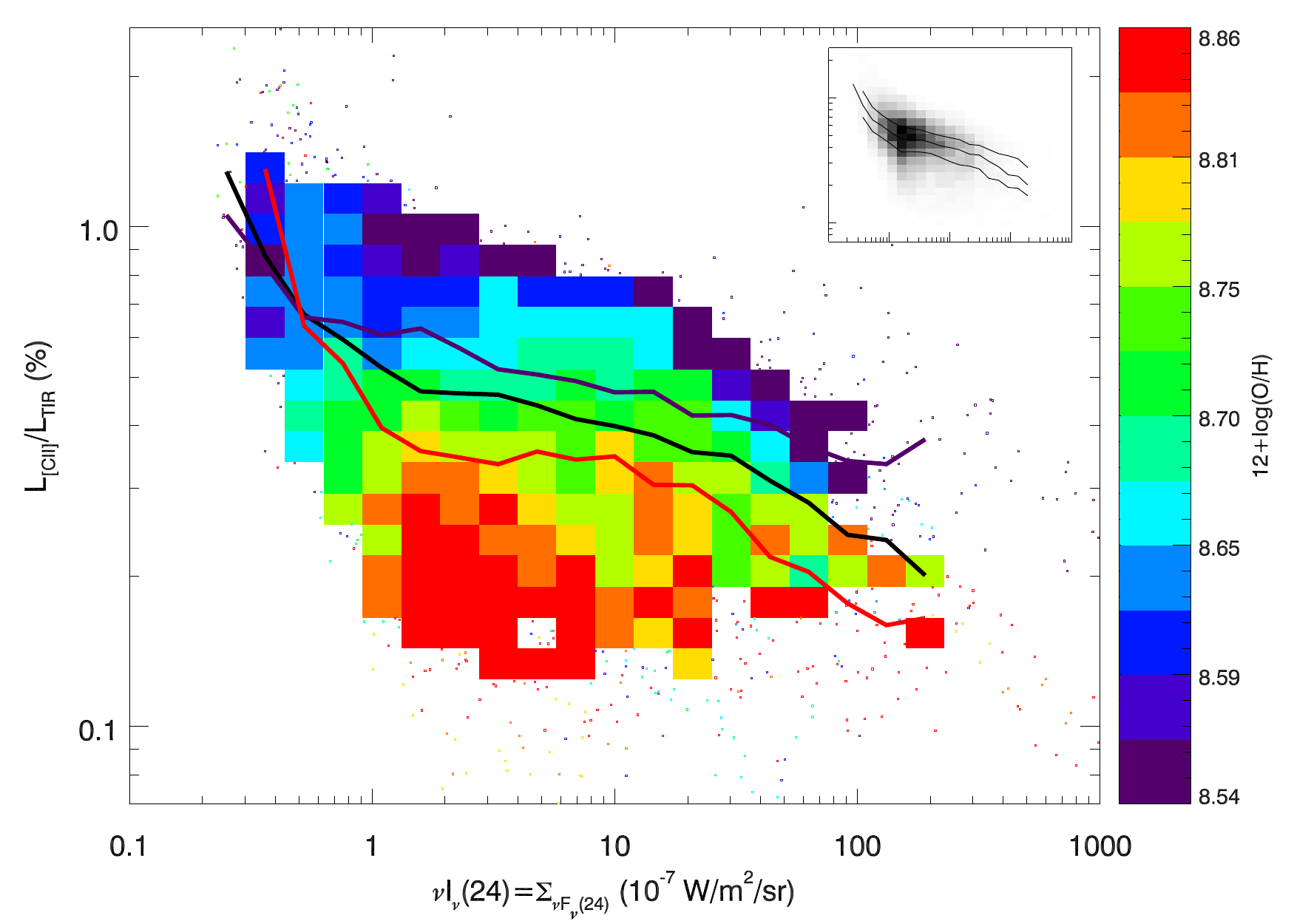}\\
\caption{The \cII\ deficit in approximately 15,000 resolved regions within \kf\ galaxies vs. the 24\um\ surface brightness $\nu I_\nu(24\um)$.   The central 750\,pc of AGN host galaxies are excluded.  Binning in $\nu I_\nu(24)$ and $L_{[\mathrm{CII}]}/L_\mathrm{TIR}$ is adopted when nine or more regions lie in a single logarithmic bin.  
A declining trend with surface brightness is evident, and is independent of the physical scale of the extraction.  Colors indicate the binned mean or individually estimated oxygen abundance, \oh, from the scale-bar at right.   The solid lines indicate the trend lines of median fractional luminosity $L_{[\mathrm{CII}]}/L_\mathrm{TIR}$ at each position in binned surface brightness \nuInu{$24\um$}.  Plotted are the overall median (black), as well as the median of those regions in the top and bottom 10\% of the \oh\ range (color-indexed to the same abundance scale; $<$10\%-ile: dark violet, $>$90\%-ile: red).  Abundance dispersion within the bins contributes to the offset between these decile trend-lines and the locus of bins with matching abundance.  In the inset at upper right, the linearly scaled density of regions per bin (418 regions maximum) is shown over the same plotting range, together with the median and the inner two quartile lines computed at each bin of surface brightness.}
\label{fig:def-abund}
\end{figure*}

Fig.~\ref{fig:def-abund} demonstrates the trends of \cIItir\ in $\sim$15,000 regions drawn from the full \kf\ sample.  Regions within 750pc of the centers of those galaxies optically classified as AGN or composite sources are excluded from this plot (but see \S\,\ref{sec:agn-impact}).  The full sample yields $\left<\cIItir\right>=0.48\pm0.21$\% by arithmetic average, and $\left<\cIItir\right>=0.46^{+0.18}_{-0.15}$\% for the median and inner 68\% of regions.  Regions are binned in logarithmic intervals of \nuInu{24} and \cIItir\ when at least 9 points fall in a bin, and are otherwise plotted individually.  The density of regions falling in each bin is shown linearly scaled in the inset (418 regions maximum).  
The full sample median at each \nuInu{24} bin (black central solid line) demonstrates the strong and monotonic drop of the fractional cooling line luminosity with 24\um\ surface brightness. The trend was found to be more significant with surface brightness than with infrared color temperatures such as \nuInu{70}/\nuInu{160} (but see \S\,\ref{sec:sfrd}).  Using an alternative measure of infrared surface brightness, \nuInu{70}, results in a similar declining trend (not shown) with nearly identical dispersion.  To test for any dependence on physical scale of the extraction region, we partitioned the region set into two subsamples --- those with physical scales above and below 0.65\,kpc (the median region size among the sample galaxies). The resulting median trend line of \cIItir\ vs. \nuInu{24} remains unchanged between these subsamples, indicating little dependency on physical extraction scale. 

At each bin of \nuInu{24} above $10^{-7}\, \mathrm{W}/\mathrm{m}^2\,\mathrm{sr}$, the standard deviation of regions with that surface brightness is approximately $\sigma(\cIItir)\simeq 0.13\%$.  Given that \cIItir\ drops to $\sim$0.2\%, this scatter is not insignificant.  Rather than being random scatter, however, a single physical parameter appears to be associated with deviation from the fiducial trend: metal abundance.

\subsection{Metallicity}

Even mild changes in metallicity can have strong effects on  radiative coupling and dust properties in the ISM of galaxies.  As metallicity is reduced, the dust-to-gas ratio drops, at first linearly with metal abundance over a wide range of metallicities \citep{2007ApJ...663..866D}, and then even faster, with growing evidence for non-linear suppression of dust abundance at metallicities below \oh$\sim$8.0 \citep{Remy-Ruyer2014,Fisher2014}, or about 1/5th the solar abundance.\footnote{Assuming \oh$_\sun$=8.69 \citep{Asplund2009}}  Perhaps not coincidentally, around this same metallicity, the fractional abundance relative to the bulk dust mass of PAH and other small grains (the principal source of photo-electrons which heat neutral gas) drops significantly \citep{Engelbracht2005,Wu2006,2007ApJ...663..866D,2007ApJ...656..770S,2008ApJ...678..804E}.  A reduced abundance of small grains would presumably \emph{decrease} photo-electric coupling and gas heating, and lower \cIItir.

On the other hand, in galaxies with lower metallicity, stars of a given mass tend to be hotter, leading to increased stellar FUV emission.  In addition, the reduced dust opacity may permit a higher fraction of UV starlight, as well as Lyman-$\alpha$ and other UV lines radiated by \HII\ regions, to penetrate deep into the neutral gas.  These effects could \emph{increase} the fraction of the energy absorbed by dust that is converted to photoelectric heating of the gas, and therefore increase the \cIItir\ ratio.

\subsubsection{The Impact of Metal Abundance on the Deficit}
To investigate the response of \cIItir\ to changes in abundance, we adopt the \oh\ abundance results of \citet{2010ApJS..190..233M}, who compiled a large number of strong-line oxygen abundance measurements, and present uniformly derived radial gradient and characteristic abundances for the SINGS sample \citep[the \textit{Spitzer}-based parent sample of \kf,][]{2003PASP..115..928K}.  Following \citeauthor{2010ApJS..190..233M}, we use the fitted radial abundance gradients, where available, to compute \oh\ in each of our regions, and then \emph{average} the two values obtained using each of the two strong line calibrations the authors chose \citep[a procedure which, for the abundance range in the sample, recovers abundances similar to those measured \emph{directly} using temperature-sensitive auroral lines, ][]{2013ApJ...777...96C}.  For the four galaxies studied not available in the \citeauthor{2010ApJS..190..233M} sample, we adopted direct or comparable strong-line abundance values.\footnote{IC0342: \citet{McCall1985}; NGC\thinspace3077: \citet{Storchi-Bergmann1994}; NGC\thinspace2146: K. Croxall, priv. comm; NGC\thinspace5457: \citet{Kennicutt2003}.  Note that \citeauthor{2010ApJS..190..233M} adopt a luminosity-metallicity calibrated abundance for galaxies without adequate line detection, affecting  $<$8\% of all regions in our sample.}  Characteristic galaxy abundances ranged from 0.17--2.85$\times$ the solar value, and the regions themselves ranged from \oh=8.46--8.88 (10\%--90\% range).

Figure~\ref{fig:def-abund} demonstrates the impact metallicity has on the fractional luminosity of \cII.  The bins (each containing 9 or more distinct regions within the sample) are colored according to the linear mean of their abundance, $12+\log(\left<\mathrm{O}/\mathrm{H}\right>)$.  Since gas heating and therefore \cII\ cooling depends critically on the coupling of UV/optical photons to small dust grains, and since both the grain-ionizing starlight and the abundance of small grains themselves are affected by varying gas metal content, it is perhaps not surprising that, at a given starlight surface density, the substantial scatter in the \cII\ deficit is driven primarily by changes in metallicity.  

Although the overall trend is towards deeper deficits with increasing surface brightness, at each surface brightness, regions with lower metallicity typically exhibit increased \cIItir, and those with higher metallicities fall to deeper deficits.  The modest range of \oh\ scaling in Fig.~\ref{fig:def-abund} reflects the limited dynamic range of abundance in the portion of the \kf\ sample covered by \cII.  At the high metallicity end, this limit is related to the chemical enrichment history of the universe --- local galaxies do not attain much more than twice solar abundance \citep{Pilyugin2007,Zahid2014}. At the low end, the dominant effect is limited coverage in the far outskirts of galaxies where metallicities are lowest \citep[e.g.][though see \S\,\ref{sec:discussion} for a discussion of deficit work in lower metallicity samples]{Bresolin2012}. 
Another impact is the abundance scatter within each bin (a median of $\sigma_{\log(\mathrm{O}/\mathrm{H})}=0.13$), which serves to compress the vertical offset between bins of different metallicity.  This per-bin scatter may stem in part from uncertainties in the measurement and calibration of abundance gradients in the sample, or from breakdowns in the assumption of purely radial abundance variation.  This effect also explains the compression towards the median of the high-metallicity (top 10\%) and low-metallicity (bottom 10\%)  trend lines in Fig.~\ref{fig:def-abund}.  
Yet even with imperfect abundances with values ranging just over a factor of two, the impact on deviations from the overall deficit trend with surface brightness are clear.

\subsubsection{Radial Dependence of the Deficit}
Several studies have identified trends of increasing \cIItir\ with radius in star forming galaxies \citep{Madden1993,Nikola2001,Kramer2013,Kapala2015}.  Since abundance is assumed to drop monotonically with radius, it is natural to question whether \oh\ is acting merely as a surrogate for some other physical property of galaxies which varies radially (aside, of course, from the radial variations in infrared surface brightness itself).  Systematic uncertainties in the determination of \oh\ gradients are also a potential concern.  To investigate these effects and the significance of the deficit dependence on metallicity, we used two approaches.  First, we suppressed all radial abundance gradients, and adopted in their place a single characteristic abundance for each sample galaxy, drawn from the strip-averaged values of \citet{2010ApJS..190..233M}.  Even when only a single characteristic abundance value per galaxy is used, the same basic pattern recurs: at a given infrared surface brightness, increasing oxygen abundance drives the deficit to deeper values.  We also directly investigated the residuals in \cIItir\ by subtracting a smoothed spline fit to the median line with 24\um\ surface brightness in Fig.~\ref{fig:def-abund}.  We found that the residuals from the overall trend with \nuInu{24} correlate significantly more strongly with \oh\ than with either physical radius or scaled radius (relative to the optical diameter along the major axis).

\section{Star Formation Rate Density Across Redshift}
\label{sec:sfrd}

\begin{figure*}
  \plotone{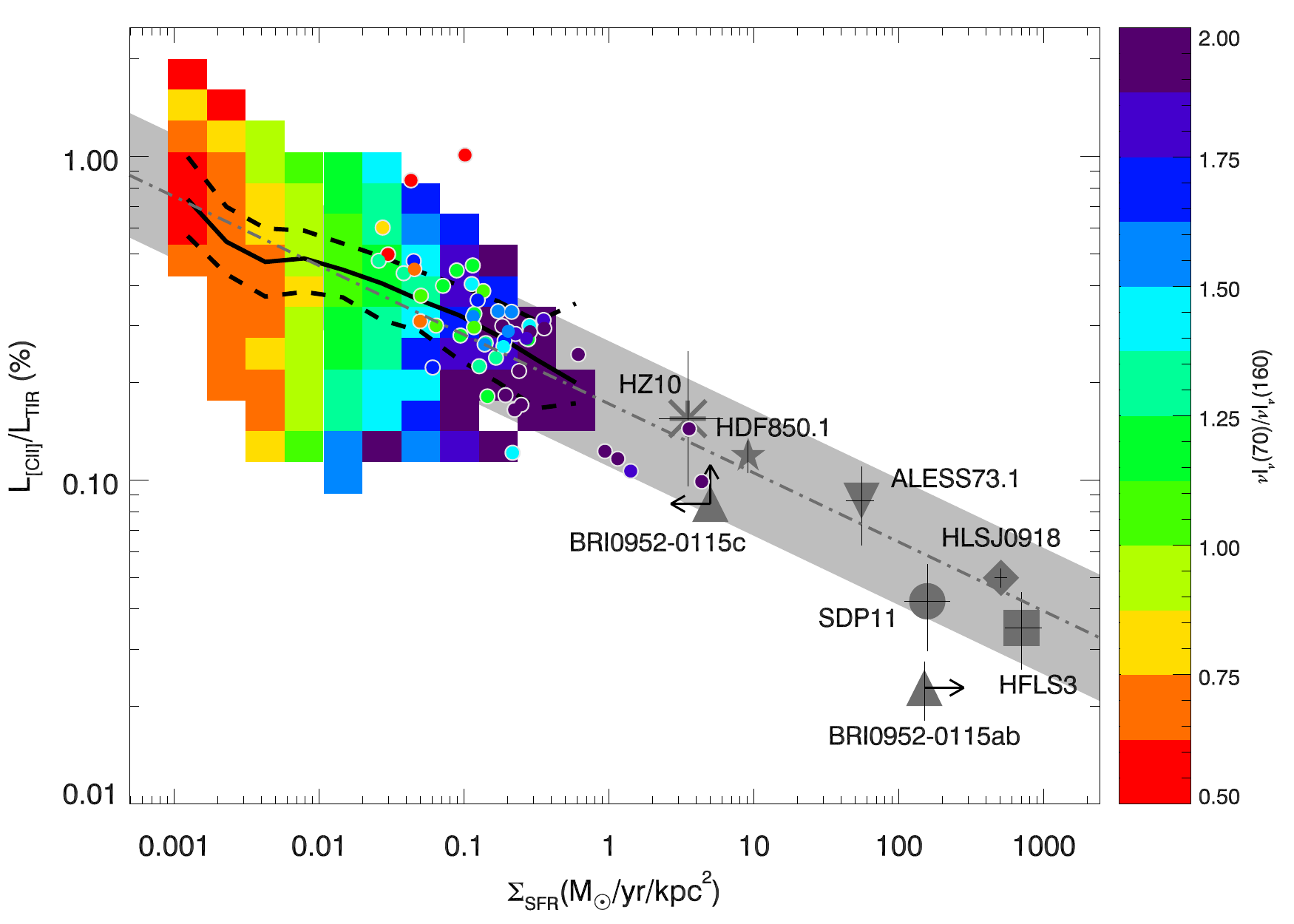}
  \caption{The trend in \cII\ deficit with star formation rate density for \kf\ regions, binned using the methodology of Fig.~\ref{fig:def-abund}.  Also included are well resolved nearby luminous infrared galaxies from the GOALS survey (filled circles), and selected high redshift sources from z=1.8--6.4 with resolved \cII\ emission (gray points).  The \kf\ and GOALS sources are color-coded by their binned median or individual dust color temperatures \nuInu{70}/\nuInu{160}, indicated on the color scale at right. The median and upper and lower quartile trend lines are shown for the \kf\ sample (solid and dashed lines, respectively).  These give a better impression of the true scatter of resolved regions about the trend, as the binning method accentuates low-density outlier regions.  A fit to the binned median, GOALS, and high-redshift points is shown as the continuous (dot-dashed) line, with the underlying fit uncertainty shaded gray.  High redshift sources from \citet{Gallerani2012}--BRI 0952-0115, z=4.4, unlabeled upward triangles; \citet{Walter2012}--HDF850.1, z=5.2; \citet{Riechers2013}--HFLS3, z=6.34; \citet{Bussmann2013} \& \citet{Ferkinhoff2014}--SDP11, z=1.8; \citet{Rawle2014}--HLSJ0918(Ra), z=5.234; \citet{DeBreuck2014}--ALESS73.1, z=4.8; and \citet{Capak2015}--HZ10, z=5.66.}
\label{fig:def-hz}
\end{figure*}

The areal density with which galaxies form new stars fundamentally impacts their physical character, altering the temperature, pressure and kinetic state of interstellar gas, the size and distribution of molecular clouds, and the efficiency of star formation.  In star-forming galaxies, on scales near or above a kiloparsec, \sfrd\ varies over a remarkably wide range, from below $10^{-5}\:\sfdunt$ in nearby low-efficiency dwarf galaxies \citep[e.g.][]{Cook2012}, up to ``maximal starbursts'' which are Eddington-limited by the impact of radiation pressure on dust at $\sfrd\simeq 1000\:\sfdunt$ \citep{Thompson2005}.  Given the strong demonstrated dependency of \cIItir\ on 24\um\ surface brightness, it is natural to consider whether \sfrd\ itself underlies this trend.

Fig.~\ref{fig:def-hz} explores the dependence of the \cII\ deficit on \sfrd.  It combines the resolved regions within \kf\ galaxies (on physical scales ranging from 0.2--1.6\,kpc) with central values from resolved local LIRGs and ULIRGs from the GOALS sample of \citet{Diaz-Santos2013}, as well as a sample of high-redshift galaxies at z=1.8--6.4 with directly measured physical size scales from well-resolved \cII\ and continuum interferometic measurements.  For the \kf\ galaxies, \sfrd\ is obtained from a hybrid indicator which combines FUV and 24\um\ surface brightness of \citet{Hao2011}.\footnote{Compared to 24\um-only calibrations, the addition of FUV photometry increases \sfrd\ only below $\sim 0.01\,\sfdunt$, and typically by less than 30\%.}   In the central GOALS regions, \sfrd\ is obtained directly from 24\um\ surface brightness using the prescription of \citet{Rieke2009}.  In both cases, measurements are performed at a matched common resolution of the PACS 160\um\ beam (11\farcs2).  Since GOALS sources are typically small compared to the 160\um\ beam, this required substantially trimming the sample to include only the 51 galaxies with well-resolved \textit{Spitzer}/MIPS 24\um\ and \textit{Spitzer}/IRS 15\um\ continuum emission \citep[measured along the slit, see][]{Diaz-Santos2013}.

The resolved \kf\ regions (colored bins, omitting low-density points as well as AGN centers, as in Fig.~\ref{fig:def-abund}) and individual GOALS galaxies are color-coded by their (mean) dust color temperature \nuInu{70\um}/\nuInu{160\um}. In the case of GOALS, this color temperature is derived from the nearby 63\um\ and 158\um\ continuum fluxes underlying the \oI\ and \cII\ lines,  using an average \citet{Dale2002} template to account for the slight wavelength offsets.  A clear trend of increasing color temperature at higher \sfrd\ is evident, although with some mixing between the two --- bins of fixed values of \nuInu{70}/\nuInu{160} ratio do not lie at a single \sfrd, and the scatter of color temperature at a given bin of \sfrd\ is significant: $\sigma({\nuInu{70}/\nuInu{160}})=0.38$.  The median and inner quartile trends of the resolved \kf\ regions are shown in solid and dashed lines. The higher star formation rate density GOALS sample clearly extends the \cIItir\ decline set by these resolved regions.

High redshift sources (see Fig.~\ref{fig:def-hz} caption) were included directly from reported \sfrd\ values or computed from sizes and IR luminosities, typically based on SED-fitting of several IR continuum measurements and utilizing the resolved physical sizes from \cII\ and nearby continuum maps.  Where necessary, conversion to TIR luminosities from FIR (42--122\um, 2.4$\times$) were made using the M82 template of \citet{Polletta2007}, and from IR (8--1000\um, 1.06$\times$) using the mean template of \citet{Dale2002}.  In all cases expected uncertainties are less than 20\% for conversion among these broad infrared luminosity bands.  The TIR luminosity was further converted to star formation rate using the updated scaling of \citet{Murphy2011}.  Source sizes were computed from the full-width at half maximum area of elliptical Gaussian fits to the \cII\ and/or resolved rest-frame sub-mm continuum intensity maps (averaging the two where both were available).

Taken together with resolved lower surface brightness \kf\ and GOALS samples, the high redshift galaxies extend a single trend of deepening \cII\ deficit over more than six orders of magnitude in \sfrd.  Of particular interest are the two galaxies which form part of the single galaxy group BRI 0952-0115 \citep[upward gray triangles in Fig.~\ref{fig:def-hz},][]{Gallerani2012}.  These two galaxies have approximately the same infrared luminosity of a few $\times10^{12}L_\sun$, but one is approximately 10$\times$ more extended, and thus, with its correspondingly lower \sfrd, has a much higher limiting \cIItir\ ratio.  

A single power-law fit to the binned \kf\ regions, GOALS galaxies, and high redshift sources is shown as the dot-dashed line in Fig.~\ref{fig:def-hz}.  The relatively tight correlation permits \emph{inverting} the relation to approximately estimate \sfrd\ directly from a measured value of the \cII\ fractional line luminosity:

\begin{equation}
  \label{eq:sfrd_cii}
  \Sigma_\textsc{sfr} =(12.7\pm 3.2)\;
  \left(\frac{L_{[\mathrm{CII}]}/L_\mathrm{TIR}}{0.1\%}\right)^{-4.7\pm0.8} \left(\frac{\mathrm{M}_\sun}{\mathrm{yr\;kpc}^2}\right)
\end{equation}

\noindent This relationship applies when young stars provide the dominant energy source, and only on scales greater than a few hundred parsecs.  The slope is similar whether the fit is performed to local sources only, or with high-redshift sources included.  At the highest physically meaningful \sfrd\ of $\sim\!1000\:$\sfdunt, \cIItir$\:\simeq 0.03$--$0.045$\%.  Since the slope is relatively steep, care must be taken to correctly account for input uncertainties in \cIItir. Even with minimal input uncertainties, the uncertainty in the relation itself contributes to over an order of magnitude range in the resulting \sfrd\ values (see gray band in Fig.~\ref{fig:def-hz}).  The steepness of the \sfrd--$L_{[\mathrm{CII}]}/L_\mathrm{TIR}$ relation also implies that in blind \cII\ surveys, lower \sfrd\ galaxies will be overrepresented compared to more compact star-forming systems.

An important advantage of Eq.~\ref{eq:sfrd_cii} is that, in addition to applying to regions within resolved galaxies, it can be used with the observed global luminosity ratio $L_{[\mathrm{CII}]}/L_\mathrm{TIR}$ to estimate the star formation surface density \sfrd\ if the \cII\ and TIR emitting size scales are similar \citep[as they are in resolved local galaxies, ][]{deBlok2016}. It is important to note that Eq.~\ref{eq:sfrd_cii} is \emph{not} a calibration for total star formation rate in a galaxy, as galaxies with widely differing SFR can have similar \sfrd.  If, however, the total star formation rate is known separately, Eq.~\ref{eq:sfrd_cii} permits estimation of the compactness of a given star-forming system.

\section{The AGN Impact}
\label{sec:agn-impact}
\begin{figure*}[t]
  \plotone{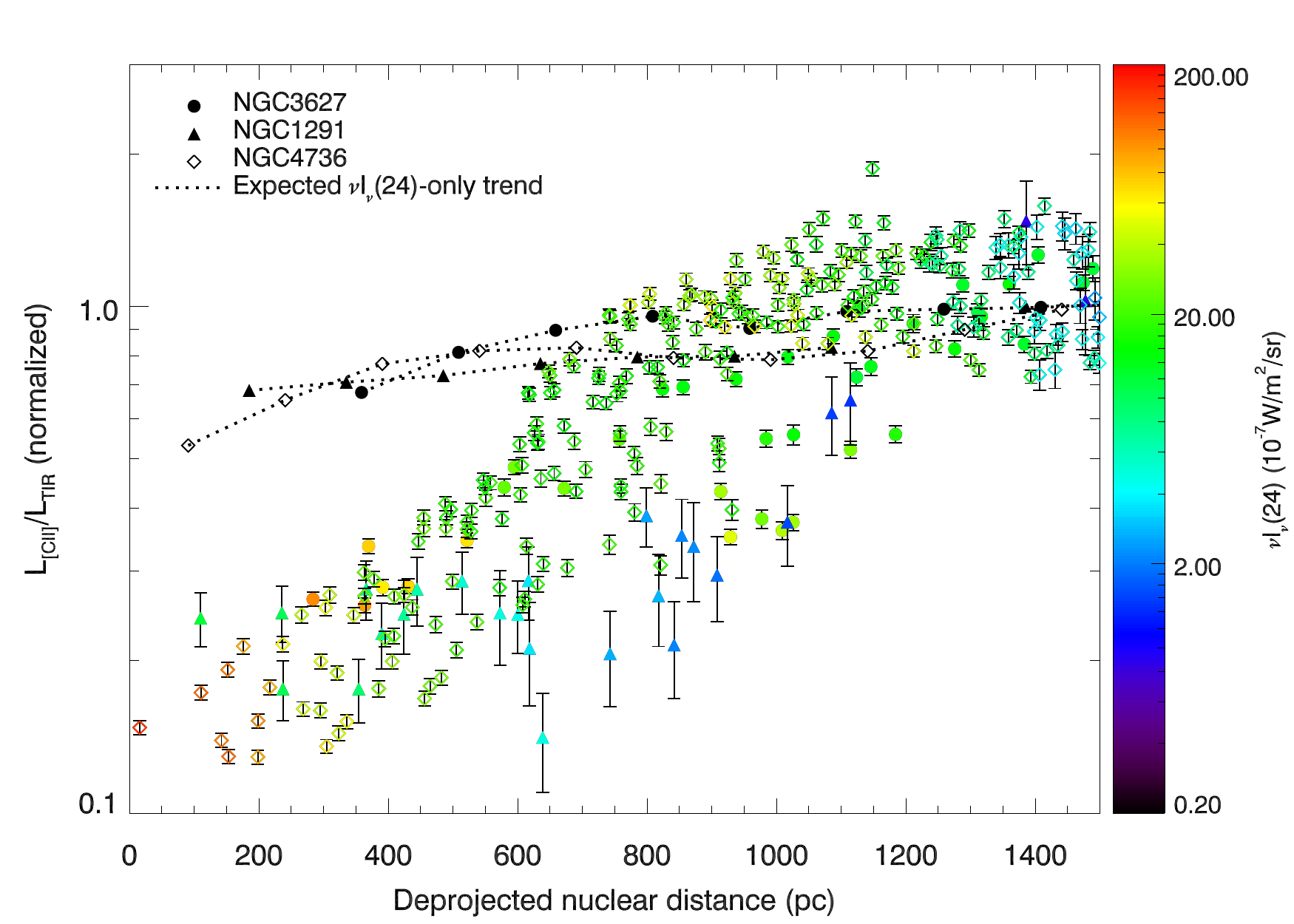}\\
\caption{The de-projected radial distribution of \cIItir\ in the inner 1.5\,kpc of three AGN systems with strong central \cII\ suppression.  In each galaxy, \cIItir\ has been normalized to its value in a 400\,pc wide annulus centered at 1.5\,kpc distance.  Individual regions are color-coded by 24\um\ surface brightness.  The \emph{expected} modest inward decline of \cIItir\ (normalized), given the increase in 24\um\ surface brightness towards the centers of each galaxy is shown by the dotted lines, computed in 150\,pc bins using the median \nuInu{24}--\cIItir\ trend line of Fig.~\ref{fig:def-abund}.  Points within 750pc of the centers of these and other AGN hosts were omitted from Figs.~\ref{fig:def-abund}~\&~\ref{fig:def-hz}.}
\label{fig:cii_agn}
\end{figure*}

AGN can reduce the relative cooling power of \cII\ in galaxies in several ways.  They are effective at producing warm central dust continuum emission (up to dust sublimation temperatures of $\sim\!\!10^3$\,K) which can substantially increase bolometric infrared luminosity and drive an apparent deficit in \cIItir.  They can also directly impact the cooling line emission itself.  One direct effect on \cII\ emission results from changes in the overall ionization state of the gas. \citet{Langer2015} modeled this effect on \cII\ emission in AGN and found substantially reduced values (to $\cIItir\sim 0.01\%$) in the inner $\sim\!10^2$\,pc of AGN hosts.  Another effect is the possible  photo-destruction of small dust grains by X-rays out to kiloparsec scales \citep{Voit1992}.  The loss of smaller grains then reduces the efficiency of photo-electric heating of neutral gas by UV photons (whether those photons are generated by stellar or accretion-driven sources).

\begin{figure*}
  \plotone{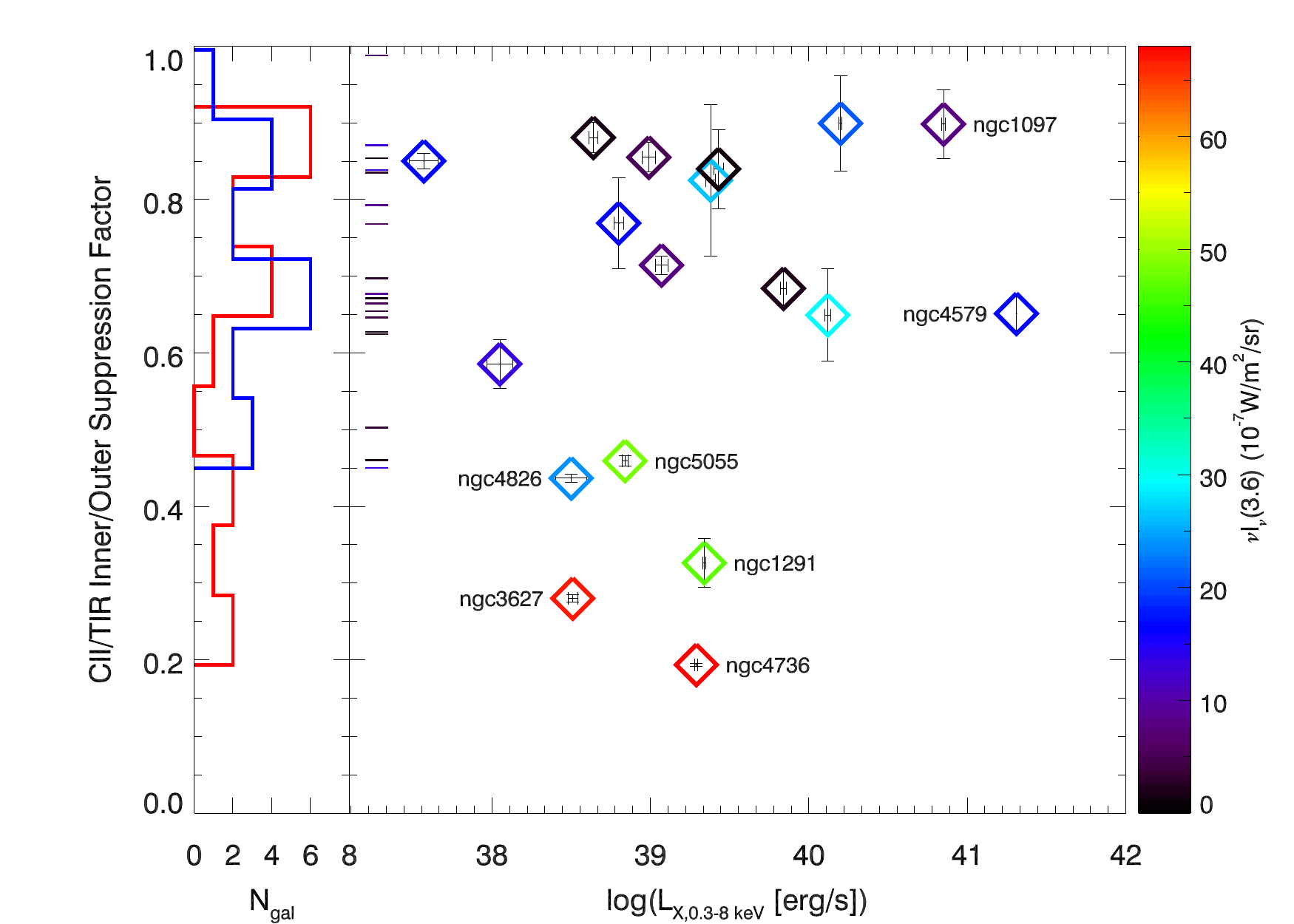}\\
\caption{The central suppression factor of \cII\ cooling power, defined as the ratio of the inner \cIItir\ value (averaged within 400\,pc of the center) to the outer value (averaged over an annulus at 1--2\,kpc). A suppression factor of 1 indicates no change in \cIItir\ from outer regions to the center.  The marginal histogram at left shows the distribution of suppression factors in AGN (red) and non-AGN (blue) hosts, and at right, as a function of central 0.3--8\,keV X-ray luminosity.  The \cII\ central suppression factors in non-AGN host centers are also shown as color-coded bars near the vertical axis.  All galaxies are color-coded by central average \nuInu{3.6\um} (a surrogate for evolved starlight intensity).}
\label{fig:cii_xray}
\end{figure*}

While \kf\ sources were not selected to host luminous AGN, approximately half of the sample, including 26 of the \cII-targeted galaxies considered here, have nuclear optical classifications consistent with AGN or composite sources \citep[either Seyfert or LINER;][]{2010ApJS..190..233M}.  Although none are bright Seyfert AGN (with typical broadband X-ray luminosities up to $10^{44}$\,erg/s), the spatial resolution available in \kf\ (down to $\sim 200$\,pc) means any AGN impact on the cooling balance can be directly resolved.  

Fig.~\ref{fig:cii_agn} illustrates the striking behavior of \cII\ in the inner 1.5\,kpc of several AGN hosts in the sample.  The steep radial trend of increasing deficit towards the centers of these three galaxies is immediately evident.  One of these, NGC\thinspace4736, is the example map shown in Fig.~\ref{fig:cii_im}, where the strong deficit of \cII\ relative to underlying infrared continuum can be readily seen in the central region, dropping in this case to $\cIItir\sim 0.05\%$.  In Fig.~\ref{fig:cii_agn}, the ratio \cIItir\ is normalized for each galaxy at a deprojected nuclear distance of $1.5\pm 0.2$\,kpc, and these three galaxies in particular show a similar pattern of rapidly declining \cIItir\ values below $\sim$1\,kpc.   While the surface brightness \nuInu{24}\ of these AGN hosts does increase towards the center (color intensity scale of the points), the relative decline in \cIItir\ for the modest range in brightness from disk to nucleus is substantially steeper than the general trend shown in Fig.~\ref{fig:def-abund}.  To demonstrate this, we have calculated the \emph{expected} change in the normalized deficit with nuclear distance due purely to changing 24\um\ surface brightness.  For each galaxy, we apply the median trend line of \cIItir\ with \nuInu{24} (Fig.~\ref{fig:def-abund}) to the average region-by-region surface brightness.  We bin radially in nuclear distance bins of 150\,pc, and normalize \cIItir\ as before at $1.5\pm 0.2$\,kpc.  These expected trends are shown in Fig.~\ref{fig:cii_agn} as the dotted black lines, which clearly demonstrate how strong the observed central suppression is in these AGN hosts, compared to the modest changes which would arise from centrally increasing surface brightness.

One more potential impact of the presence of an AGN on the heating/cooling balance is entirely indirect: especially at low fractional AGN luminosity, the optical diagnostics used to classify nuclear regions are diluted in the presence of star formation.  For this reason, most local low-luminosity AGN galaxies are dominated by older stellar populations in their centers \citep[e.g.][]{Singh2013}.  
In this interpretation, the AGN is relevant only by association with the high starlight intensity and softer spectrum which characterize their host's old central stellar population.

\citeauthor{Langer2015} predicted deviations in \cIItir\ above X-ray luminosities of $10^{43}$\,erg/s.  To investigate trends with AGN luminosity, we draw on the central 0.3--8\,keV luminosities from Chandra which \citet{Grier2011} compiled for the SINGS sample. Figure ~\ref{fig:cii_xray} shows the central suppression of \cIItir, defined as the ratio of this value in the central 400\,pc\footnote{or, the central pointing for the few galaxies with beam sizes larger than 400\,pc} relative to its average value at 1--2\,kpc.  It is immediately evident that, while all AGN host galaxies show some suppression of \cIItir\ towards their centers, in most cases this suppression is modest, approximately 30\%.  The strength of \cII\ suppression does not correlate with central X-ray luminosity, and for most AGN, it is similar to non-AGN hosts in the sample --- 30$\pm$15\% (seen as bars at left and in the marginal histogram).  But a small number of AGN sources (including the three with radial profiles shown in Fig.~\ref{fig:cii_agn}) show substantially higher central suppression factors, with \cIItir\ reduced by up to a factor of 5 from disk to center. The five most suppressed sources have 0.3--8\,keV luminosities well below the average ($L_{0.3-8\,\mathrm{keV}}=10^{40.3}$\,erg/s).  Moreover, these highly suppressed sources are in fact those with the largest central surface densities of evolved starlight, as measured by \nuInu{3.6\um} (color scale of the points).  NGC\thinspace4736, the low-luminosity AGN host with the strongest central \cIItir\ suppression, has the highest central 3.6\um\ surface brightness of the entire sample.  This suggests that the high density of soft starlight may drive the enhanced central \cIItir\ suppressions in AGN hosts.

\section{Discussion}
\label{sec:discussion}

Relative to integrated infrared luminosity, the fractional power emitted in \cII\ --- the dominant neutral gas coolant and most luminous emission line in most galaxies --- ranges over more than two orders of magnitude, from $\cIItir\gtrsim$1\% to $\lesssim$0.01\%.  This  ``cooling line deficit'' has been well studied globally in galaxies ranging from normal star-forming to ultraluminous systems, with the physical origin as yet unresolved. Such global deficits have been invoked as evidence in support of distinct physical conditions controlling the star formation process in compact luminous galaxies, vs. ``main sequence'' galaxies forming stars at low surface densities.  

A number of physical explanations for the large variations in the power of the dominant cooling line relative to dust-reprocessed emission among galaxies has been considered, including:
\begin{itemize}
\item \cII\ self-absorption, which could lead to absorbed velocity-resolved line profiles in deep deficit sightlines.

\item Strong continuum extinction at 158\um.

\item Significant and varying contributions of \cII\ from (hydrogen-) ionized gas \citep[e.g.][]{Luhman2003}.

\item Collisional quenching of \cII\ emission at high density, which would result in the dense gas coolant \oI\ compensating \citep[e.g.][]{Brauher2008}, although this effect was not seen in two \kf\ galaxies \citep{2012ApJ...747...81C}.

\item \HII\ regions with high ionization parameter, in which dust absorbs non-ionizing UV photons before they can enter neutral gas.  This should result in deficits in all cooling lines for neutral gas \citep[see][]{Gracia-Carpio2011}.

\item Small grain destruction or charging \citep[e.g.][]{Malhotra2001}, which reduces photo-electric yield and should result in correlations between \cIItir\ or the \cII/PAH ratio and, e.g., $q_\mathrm{PAH}$ \citep[the fractional mass in PAH grains,][]{Draine2007}, or PAH band ratios indicative of ionization such as 7.7\um/11.3\um\ \citep[as found by][]{2012ApJ...747...81C}.

\item The collisional coupling of gas and dust grains at high density which can ``short-circuit'' the heating/cooling balance via continuum cooling at high gas densities \citep{Spaans2000}.

\item AGN impacts on the ionization state of gas \citep[see \S\,\ref{sec:agn-impact}, and][]{Langer2015}, which would lead to a correlation between deficits and central X-ray luminosity.

\item An infrared excess relative to the rate radiation couples to the gas, for example in energetically dominant AGN/Quasars, at extreme dust column densities, or in the regions of high starlight intensities that are dominated by softer spectra from old stellar populations in central galaxy bulges \citep[e.g.][]{Groves2012,Draine2014}.

\item Other cooling channels such as rotational CO emission dominating the gas-phase cooling power, for example at low
ratios of UV field to gas density, where the atomic gas layer becomes thin and the \cII\ line intensity diminishes \citep[see][]{Wolfire1989,Kaufman1999}.  CO does not, however, appear to provide substantial global cooling in ULIRGs \citep{Rosenberg2015}.
\end{itemize}

Whatever combination of the above physical effects drives the deficit, here we have demonstrated that the same varying deficit of cooling power operates \textit{within galaxies} on scales as small as 200\,pc.  Remarkably, a single physical property controls the bulk of this variation over six orders of magnitude: the star formation rate density, with increasing \sfrd\ driving down the fractional \cII\ cooling power.  

Due to its high intrinsic luminosity, and accessibility at high-redshift \citep[e.g.][]{Carilli2013}, there is considerable interest in employing \cII\ emission as a direct tracer of star formation. The strong deficit dependence on \sfrd\ has ramifications on the use of \cII\ as a star formation indicator.  Resolved studies which explore \sfrd-$I_{[\mathrm{CII}]}$ surface brightness relationships effectively ``calibrate out'' this dependence and as a result can yield relatively unbiased indicators \citep[e.g.][who found that the \cII--SFR relation is better behaved in surface density than luminosity space]{Herrera-Camus2015}.  Applying these resolved \cII-SFR calibrations to \emph{global} luminosities is more uncertain, unless the sample exhibits a modest range of global \sfrd\ similar to the galaxies used for the calibration.  This fact is easily understood when considering the extreme range of \cII/FIR found at fixed FIR luminosity \citep[$\gtrsim250\times$,][]{Brisbin2015}.  The strong trend seen in Fig.~\ref{fig:def-abund} also explains the residual dependence of global \cII/SFR on surface brightness and dust color temperatures \citep[e.g.][]{DeLooze2014,Herrera-Camus2015}.  Even the definition of \sfrd\ itself requires some care, as different observational size scales for extended star-forming disks can yield appreciable differences.  The results here apply on size scales above 200\,pc and within the main \cII-emitting disk of galaxies.

At a given surface brightness, deficit variations are found which are not random, but are correlated with the gas-phase oxygen abundance, with decreasing metallicity associated with increasing \cIItir.  These variations are consistent with the large measured fractional luminosities of \cII\ in some very low metallicity objects \citep[e.g.][]{Hunter2001,Cormier2010,Cigan2016}.  In contrast, \citet{Cormier2015} found little dependence of \cIItir\ on \oh\ in global measurements of a large sample of low metallicity dwarf galaxies.  Since the \sfrd\ dependence is stronger than the impact of metallicity, unresolved global measures sampling widely varying local radiation field strengths might wash out these residual trends.  \citet{Vallini2015} modeled \cII\ emission at high redshift, and found a second order dependence of the predicted \cII-SFR relation with metallicity $Z$ which has the \emph{opposite} sense of our observations (at a fixed SFR, $L_{[\mathrm{CII}]}$ increases in their models as metallicity increases), but this offset may arise principally from the adopted correlation between baryonic overdensity and $Z$.

The apparent cooling line excess with decreasing metallicity at a given surface brightness (Fig.~\ref{fig:def-abund}) is difficult to reconcile with the plummeting relative abundance of small dust grains at low metallicity.  Since PAHs and related small grains provide the bulk of photo-electric heating, finding a cooling line \emph{excess} relative to large grain heating near the metallicity \oh$\:\simeq\:$8.2 where PAHs appear to begin disappearing \citep{Engelbracht2005,Wu2006,2007ApJ...663..866D,2007ApJ...656..770S,2008ApJ...678..804E} indicates that some other process must contribute.  One possibility is that the contribution of \cII\ from ionized gas simultaneously increases as metallicity drops.  This has been investigated in detail by Croxall et al. (in prep) using \nII\ 205\um\ emission (which shares \cII's critical density in ionized gas and thus tracks it closely).  They find that only a modest fraction ($<\sim 40\%$) of \cII\ originates in ionized gas and that this fraction in fact \emph{decreases} significantly as metallicity drops, essentially compounding the problem.  It may also be the case that other small grains with abundances which are insensitive to declining metallicity and for which PAH grains do not act as observational surrogates can still contribute substantial heating \citep[e.g.][]{Galliano2005,Israel2011}. 
Cosmic rays may also play a role, if their heating contribution (relative to dust photoelectric heating) increases at low metallicity.  This could occur, for example, due to (1) an increasing fraction of energy going into cosmic rays as a result of reduced radiative cooling in supernova blastwaves, and (2) reduced absorption of FUV photons (and consequent photoelectric heating) in low metallicity galaxies with lower dust abundances.
A final explanation is that \cII\ emission at low metallicity is dominated by larger ``\cII-regions'' surrounding hydrogen ionization zones, made possible by the increased penetration depth of photons with $h\nu\lesssim 13.6$\,eV in the reduced dust environments, which can maintain a larger fraction of the ISM's carbon as C$^+$, and/or increase the grain heating efficiency due to a more dilute radiation field heating the gas \citep[e.g.][]{Poglitsch1995,Madden1997,Israel2011}. A similar argument based purely on scaling radiation density and resulting cloud structure is given by \citet{Narayanan2016}.  In strong recombination regions, the abundance of C$^+$ could also be enhanced by the reduced rate of carbon recombination on grain surfaces expected as the small grain abundance drops \citep{Kaufman1999}.  Further discriminating among the physical mechanisms underlying the deficit's response to metallicity may require partitioning \cII\ emission into its ionized, neutral, dense and diffuse emitting environments.

Some studies have suggested that the \cII\ deficit arises solely from the impact of AGN on the cooling balance in the ISM \citep[e.g.][but see \citealt{Ibar2015} for an opposing view]{Sargsyan2012}. The continuity of the presented deficit trend from hyperluminous high redshift galaxies down to sub-kpc scales within normal star-forming galaxies demonstrates this not to be the case generally.  While there is a strong central deficit impact of some AGN in the \kf\ sample, as discussed in \S\,\ref{sec:agn-impact}, it is likely indirect, related to the high density of low energy starlight photons which accompany lower luminosity AGN.  As modeled by \citet{Draine2014} for M31's bulge, the softening central starlight spectrum couples less efficiently to the PAHs than to the bulk grain populations.  This presumably would lead to reduced photo-electric heating efficiency of the gas, as less photo-ionization energy per photon is available when average photon energy drops.  A detailed model of this effect with varying starlight spectra could provide a valuable test of this scenario.  

At low luminosity, a related issue could be the recently revealed ambiguous power source of many LINER galaxies \citep[with resolved studies showing extended LINER-like emission inconsistent with black hole accretion,][]{Singh2013,Belfiore2016}.  The optical classification scheme of \citet{2010ApJS..190..233M} we utilize did not separate Seyfert and LINER types.  Whether other expected direct impacts of AGN, for example on the kpc-scale carbon ionization state or the wide-scale small grain population, can be disentangled from the effects of changing starlight conditions in the centers of AGN hosts is unknown, but resolved \cII\ investigations in AGN with X-ray luminosities above $10^{43}$\,erg/s would be illuminating.

At the highest luminosities, the dust continuum of \cII-emitting galaxies can be dominated by powerful AGN, rendering Eq.~\ref{eq:sfrd_cii} inapplicable.  An example of this is a galaxy at z=4.6 with the highest currently measured infrared luminosity \citep[$L_\mathrm{TIR}\sim 2.2\times 10^{14} L_\sun$,][]{Diaz-Santos2016}.   Like other high luminosity systems which host extreme AGN, it has a resolved continuum size much smaller than its \cII\ disk, and a remarkably deep \cII\ deficit ($\cIItir\sim0.003\%$).  For such a deep deficit, Eq.~\ref{eq:sfrd_cii} would imply an unphysical star formation rate density in excess of $10^7\,\sfdunt$.  Even using a star-forming SED model to compute a reduced $L_\mathrm{TIR}$ by scaling from the much lower far-infrared (42-122\um) luminosity, which is often argued to suffer less contamination from AGN-heated dust than the bolometric infrared, leads to $\sfrd \gtrsim 10^5\,\sfdunt$.  Very deep deficits and the resulting unrealistically high \sfrd's implied by Eq.~\ref{eq:sfrd_cii} likely indicate the presence of dominant AGN heating.

\section{Conclusions}
\label{sec:conclusions}

We have explored \cII\ emission in approximately 15,000 regions within 54 \kf\ galaxies,  supplementing these measurements with well-resolved \cII\ observations of nearby luminous infrared galaxies and interferometrically resolved high-redshift \cII\ emitters.  The uncovered physical trends in the deficit connect environments ranging from low surface brightness regions in moderate star-forming systems to extreme starbursts at the theoretical limits of star formation surface density. We find:

\begin{enumerate}
\item The \cII\ deficit, in which \cIItir\ exhibits large variations contrary to the simple expectations of heating/cooling balance, is present within galaxies down to scales of 200\,pc.  This strongly implies that whatever underlying process(es) drive the deficit, they are local physical processes of interstellar gas not related to global galaxy properties like bulk luminosity or presence on the galaxy main-sequence.

\item Within normal star-forming galaxies, \cIItir\ declines markedly from over 1\% down to $\sim$0.1\% as surface brightness \nuInu{24\um} increases, with an average value $\left<\cIItir\right>=0.48\pm0.21$\%.

\item At fixed surface brightness, the gas phase metallicity within galaxies, \oh, correlates with the residual variation in \cIItir, with reduced metal abundance associated with higher \cIItir\ values (i.e., reducing the deficit), and increased metallicity associated with deeper deficits.  Given the apparent declining abundance with falling metallicity of the small grains thought to dominate photoelectric heating of the gas, this is a surprising result.

\item The variation of the deficit with 24\um\ and 70\um\ surface brightness can be most directly interpreted as a trend with \sfrd.  When combining resolved \cII\ measurements of luminous infrared and high redshift galaxies from z=1.8--6.4, the trend found in nearby galaxies smoothly extends over more than six orders of magnitude down to the maximal starburst at $\sfrd\!\sim\!1000\:\sfdunt$ and $\cIItir\sim0.03$\%.

\item By fitting and inverting the relation between \cIItir\ and \sfrd, the approximate star formation surface density can be estimated using resolved \emph{or unresolved} measurements of the fractional \cII\ luminosity, with the deepest \cII\ deficits corresponding to the highest densities of star formation.

\item Unexpectedly large deficits in the resolved \kf\ sample occur in the centers of several galaxies hosting low-luminosity AGN, with a very steep radial suppression of \cIItir\ inwards from de-projected distances of 1.5kpc to the center of up to a factor of 5.

\item While all sample galaxies, including AGN hosts, exhibit deeper deficits in their centers, the typical suppression of \cIItir\ is modest: $\sim$30\% below the 1.5\,kpc inner disk average.  Those AGN host galaxies with substantially greater central depressions do \emph{not} host the most luminous AGN (in terms of 0.3--8keV X-ray luminosity), but instead harbor central bulges with the highest surface brightness of starlight at 3.6\um.  This can be explained if the high intensity but softer starlight from the old stellar populations heating the dust in the centers of many AGN hosts drives continuum emission with reduced radiative coupling to the gas.

\item Galaxies with deficits considerably deeper than \cIItir$\sim$0.01\% imply unphysical star formation rate densities well above several thousand \sfdunt --- a potential indicator of dominant AGN contribution to the infrared luminosity.

\end{enumerate}

\acknowledgments 

This work is based in part on observations made with \textit{Herschel}, a
European Space Agency Cornerstone Mission with significant participation
by NASA. Support for this work was provided by NASA through an award
issued by JPL/Caltech.   We thank Steve Hailey-Dunsheath, T. Rawle, and Tanio Diaz-Santos for advanced access to their compiled \cII\ data sets.  We also thank them, as well as Gordon Stacey,  Carl Ferkinhoff, M. Kapala, and R. Decarli, for helpful discussions which improved this work.  JDS gratefully acknowledges visiting support from the Alexander von Humboldt Foundation and the Max Planck Institute f\"{u}r Astronomie as well as support from the Research Corporation for Science Advancement through its Cottrell Scholars program.

\bibliographystyle{apj}
\bibliography{}

\end{document}